\documentstyle[11pt]{article}
\begin{document}
\begin{titlepage}
\hfill{UQMATH-93-04}
\vskip.3in
\begin{center}
{\huge Casimir Invariants for Quantized Affine Lie Algebras}
\vskip.3in
{\Large M.D.Gould} and {\Large Y.-Z.Zhang}
\vskip.3in
{\large Department of Mathematics, University of Queensland, Brisbane,
Qld 4072, Australia}
\end{center}
\vskip.6in
\begin{center}
{\bf Abstract:}
\end{center}
Casimir invariants for quantized affine Lie algebras are constructed and their
eigenvalues computed in any irreducible highest weight representation.

\end{titlepage}
\noindent
Casimir invariants for quantum (super)groups\cite{Drinfeld}\cite{Jimbo} have
been studied by a number of authors \cite{Reshetikhin}\cite{GZB}
\cite{Lee}  and general methods
for constructing these invariants are proposed
\cite{GZB}\cite{Lee}. The aim of this short letter is to apply the method
to quantized affine Lie algebras and to obtain the Casimir invariants
for the case at hand.

Quantized affine Lie algebras are defined as $q$-deformation of classical
(universal enveloping)
affine Lie algebras with a symmetrizable, generalized Cartan matrix \cite{Kac}.
To begin with, let $A=(a_{ij})_{0\leq i,j\leq r}$ be a symmetrizable,
generalized Cartan matrix in the sense of Kac\cite{Kac}.
Let $\hat{\cal G}$ denote the affine Lie algebra
associated with the corresponding symmetric Cartan
matrix $A_{\rm sym}=(a^{\rm sym}_{ij})=(\alpha_i,\alpha_j),~~i,j=0,1,...r$
\,,~$r$ is the rank of the corresponding finite-dimensional simple Lie algebra
${\cal G}$. Then the quantum algebra $U_q(\hat{\cal G})$ is defined by
generators: $\{e_i,~f_i,~q^{h_i}~(i=0,1,...,r),~q^d\}$ and relations
\begin{eqnarray}
&&q^h.q^{h'}=q^{h+h'}~~~~(h,~ h'=h_i~ (i=0,1,...,r),~d)\nonumber\\
&&q^he_iq^{-h}=q^{(h,\alpha_i)} e_i\,,~~q^hf_iq^{-h}=q^{-(h,\alpha_i)}
  f_i\nonumber\\
&&[e_i, f_j]=\delta_{ij}\frac{q^{h_i}-q^{-h_i}}{q-q^{-1}}\nonumber\\
&&\sum^{1-a_{ij}}_{k=0}(-1)^k e_i^{(1-a_{ij}-k)}e_je_i^{(k)}
   =0~~(i\neq j)\nonumber\\
&&\sum^{1-a_{ij}}_{k=0}(-1)^k f_i^{(1-a_{ij}-k)}f_jf_i^{(k)}
   =0~~(i\neq j)\label{relations1}
\end{eqnarray}
where
\begin{equation}
e_i^{(k)}=\frac{e^k_i}{[k]_q!},~~~f^{(k)}_i=\frac{f^k_i}{[k]_q!}
\,,~~~[k]_q=\frac{q^k-q^{-k}}{q-q^{-1}}\,,~~~[k]_q!=[k]_q[k-1]_q\cdots [1]_q
\end{equation}

The Cartan subalgebra (CSA) of $\hat{\cal G}$ is generated by $\{h_i,\;i=0,1,
\cdots,r\,;\,d\}$. However, we will choose as the CSA of $\hat{\cal G}$
\begin{equation}
{\cal H}={\cal H}_0\bigoplus{\bf C}\,c\bigoplus{\bf C}\,d
\end{equation}
where $c=h_0+h_\psi$,~ $\psi$ is the highest root of ${\cal G}$ and
${\cal H}_0$ is a CSA of ${\cal G}$.

The algebra $U_q(\hat{\cal G})$ is a Hopf algebra with coproduct, counit and
antipode similar to the case of $U_q(\cal G)$:
\begin{eqnarray}
&&\Delta(q^h)=q^h\otimes q^h\,,~~~h=h_i,~d\,,~~~i=0,1,\cdots, r\nonumber\\
&&\Delta(e_i)=q^{-h_i/2}\otimes e_i+e_i\otimes q^{h_i/2}\nonumber\\
&&\Delta(f_i)=
q^{-h_i/2}\otimes f_i+f_i\otimes q^{h_i/2}\nonumber\\
&&S(a)=-q^{h_\rho}aq^{-h_\rho}\,,~~~a=e_i,f_i,h_i,d\label{coproduct1}
\end{eqnarray}
where $\rho$ is the half-sum of the positive roots of $\hat{\cal G}$.
We have omitted the formula for counit since we do not need them.

Let $\Delta'$ be the opposite coproduct: $\Delta'=T\Delta$, where $T$ is
the twist map: $T(x\otimes y)=y\otimes x\,,~\forall x,y\in U_q(\hat{\cal G})$.
Then $\Delta$ and $\Delta'$ is related by the universal $R$-matrix $R$
in $U_q(\hat{\cal G})\otimes U_q(\hat{\cal G})$ satisfying, among others,
\begin{eqnarray}
&&\Delta'(x)R=R\Delta(x)\,,~~~~~\forall x\in U_q(\hat{\cal G})\nonumber\\
&&R^{-1}=(S\otimes I)R\,,~~~~~R=(S\otimes S)R
\end{eqnarray}

The representation theory of $U_q(\hat{\cal G})$ bears much similarity to
that of $\hat{\cal G}$ \cite{Rosso}\cite{ZG}. In particular,
classical and corresponding quantum representations have the same dimension and
weight spectrum. Following the usual convention, we denote the weight of
a representation by $\Lambda\equiv (\lambda,\kappa,\tau)$, where $\lambda$
$\in {\cal H}_0^*\subset{\cal H}^*$ is a weight of ${\cal G}$ and
$\kappa=\Lambda(c)\,,\,\tau=\Lambda(d)$. The non-degenerate form $(~,~)$
on ${\cal H}^*$ is defined by \cite{GO}
\begin{equation}
(\Lambda,\Lambda')=(\lambda,\lambda')+\kappa\,\tau'+\kappa'\,\tau
\,,~~~~{\rm for}~\Lambda'\equiv (\lambda',\kappa',\tau')
\end{equation}
With these notations we have
\begin{equation}
\rho=(\rho_0,0,0)+g(0,1,0)
\end{equation}
where $\rho_0$ is the half sum of positive roots of ${\cal G}$ and
$2g=(\psi,\psi+2\rho_0)$.

We will call
\begin{equation}
D_q[\Lambda]={\rm tr}(\pi_{\Lambda}(q^{2h_\rho}))
\end{equation}
the $q$-dimension of the integrable irreducible highest weight representation
$\pi_\Lambda$: explicitly \cite{Kac}
\begin{eqnarray}
&&D_q[(\lambda,\kappa,\tau)]=q^{2g\tau}\,\bar{D}_q[(\lambda,\kappa,0)]
 \,, \nonumber\\
&&\bar{D}_q[(\lambda,\kappa,0)]=D^0_q[\lambda]\prod_{t=1}^\infty
\left (\frac{1-q^{-2t(\kappa+g)}}{1-q^{-2tg}}\right )^r
\prod_{\alpha\in \Phi_0}\prod_{t=1}^\infty\frac{1-q^{-2(\lambda+\rho_0,\alpha)
-2t(\kappa+g)}}{1-q^{-2(\rho_0,\alpha)-2tg}}\label{q-dimension1}
\end{eqnarray}
with $D^0_q[\lambda]$ given by
\begin{equation}
D^0_q[\lambda]=\prod_{\alpha\in\Phi^+_0}\frac{q^{(\lambda+\rho_0,\alpha)}
-q^{-(\lambda+\rho_0,\alpha)}}{q^{(\rho_0,\alpha)}-q^{-(\rho_0,\alpha)}}
\end{equation}
where $\Phi_0$ and $\Phi^+_0$ denote the set of roots and positive roots of
${\cal G}$, respectively. Note that the
$q$-dimension (\ref{q-dimension1}) is absolutely covergent for $|q|\,>\,1$.

With the coproduct (\ref{coproduct1}), the universal $R$-matrix has the general
form
\begin{equation}
R=\sum_ta_t\otimes b_t\label{99}
\end{equation}
where $\{a_t\,|\,t=1,2,\cdots\,\}$ and $\{b_t\,|\,t=1,2,\cdots\,\}$ are
basis of subalgebras $U_q^\pm(\hat{\cal G})$ of $U_q(\hat{\cal G})$,
generated by $\{e_i,h_i~(i=0,1,\cdots, r);d\}$
and $\{f_i,h_i~(i=0,1,\cdots,r);d\}$, respectively. Then according to
\cite{Drinfeld}\cite{FR}, there exists a distinguished element associated
with (\ref{99})
\begin{equation}
u=\sum_tS(b_t)a_t\label{u}\;,
\end{equation}
which has inverse
\begin{equation}
u^{-1}=\sum_tS^{-2}(b_t)a_t\label{u-1}
\end{equation}
and satisfies
\begin{eqnarray}
&&S^2(a)=uau^{-1}\,,~~~\forall a\in U_q(\hat{\cal G})\nonumber\\
&&\Delta(u)=(u\otimes u)(R^TR)^{-1}
\end{eqnarray}
where $R^T=T(R)$.  One can show that
$v=uq^{-2h_\rho}$ belongs to the center
of $U_q(\hat{\cal G})$ and satisfies
\begin{equation}
\Delta(v)=(v\otimes v) (R^TR)^{-1}\label{vv1}
\end{equation}
Moreover, on an integrable irreducible representation of highest weight
$\Lambda\equiv (\lambda,\kappa,\tau)\in D^+$,~
$D^+$ denotes the set of all dominant integral weights,
the Casimir operator $v$ takes the eigenvalue \cite{ZG}
\begin{equation}
\chi_\Lambda=q^{-(\Lambda,\Lambda+2\rho)}
\end{equation}
The following result is proven in \cite{GZB}
\vskip.1in
\noindent {\bf Proposition 1:} Let $V(\Lambda)$ be an irreducible highest
weight $U_q(\hat{\cal G})$-module with highest weight $\Lambda\in D^+$.
If the operator $\Gamma\in U_q(\hat{\cal G})\otimes {\rm End}
V(\Lambda)$ satisfies $\Delta_\Lambda(a)\Gamma=\Gamma\Delta_\Lambda(a)$,~~
$\forall a\in U_q(\hat{\cal G})$, where $\Delta_\Lambda=(I\otimes \pi_\Lambda)
\Delta$, then
\begin{equation}
C=(I\otimes {\rm tr})\{[I\otimes \pi_\Lambda(q^{2h_\rho})]\Gamma\}
\end{equation}
belongs to the center of $U_q(\hat{\cal G})$, i.e. $C$ is a Casimir invariant
of $U_q(\hat{\cal G})$.

We note that for
\begin{equation}
\Gamma=(I\otimes\pi_{\Lambda_0})R^TR\label{gamma}
\end{equation}
we have
$\Delta_{\Lambda_0}(a)\Gamma=\Gamma\Delta_{\Lambda_0}(a)\,~\forall a\in
U_q(\hat{\cal G})$. Therefore, from proposition 1,
\begin{equation}
C^{\Lambda_0}=(I\otimes{\rm tr})\{[I\otimes\pi_{\Lambda_0}(q^{2h_\rho})]
\Gamma\}=\sum_{s,t}{\rm tr}(\pi_{\Lambda_0}(q^{2h_\rho}a_sb_t))b_sa_t
\end{equation}
is a Casimir invariant. To compute its eigenvalue on
irreducible highest weight $U_q(\hat{\cal G})$-module $V(\Lambda)$ with
highest weight $\Lambda$, we consider
\begin{equation}
\chi_\Lambda(C^{\Lambda_0})=<\Lambda|C^{\Lambda_0}|\Lambda>=\sum_{s,t}<\Lambda|
{\rm tr}(\pi_{\Lambda_0}(q^{2h_\rho}a_sb_t))b_sa_t|\Lambda>\label{1}
\end{equation}
where $|\Lambda>$ stands for the highest weight vector of highest weight
$\Lambda$. We immediately see that
only these $a_t\,,\,b_t$ made up entirely of Cartan elements
of $U_q(\hat{\cal G})$ contribute. The basis for such elements of
$U_q^-(\hat{\cal G})$ and $U_q^+(\hat{\cal G})$
are given by respectively,
\begin{equation}
\prod_{i=1}^r\;\frac{(H_i)^{m_i}}{(m_i)!}\frac{c^{m_c}}{(m_c)!}
\frac{d^{m_d}}{(m_d)!}
\end{equation}
and
\begin{equation}
\prod_{i=1}^r\,(H^i{\rm ln}q)^{m_i}\,(d\,{\rm ln}q)^{m_c}\,(c\,{\rm ln}q)^{m_d}
\end{equation}
where $m_i\;,\;m_c\;,\;m_d\;\in\,{\bf Z}^+$ and $\{H^i\}$ and $\{H_i\}$
are defined by
\begin{equation}
\sum_{i=1}^r\,\Lambda(H_i)\Lambda'(H^i)=(\lambda,\lambda')\,,~~~~
\forall \Lambda=(\lambda,\kappa,\tau)\,,\,\Lambda'=(\lambda',\kappa',\tau')
\,\in\,{\cal H}^*
\end{equation}
Therefore, (\ref{1}) takes the form
\begin{eqnarray}
\chi_{\Lambda}(C^{\Lambda_0})&=&\sum_{{\bf m},{\bf l}}\;<\lambda|
\frac{(H_1)^{m_1}}{(m_1)!}\cdots\frac{(H_r)^{m_r}}{(m_r)!}\frac{c^{m_c}}{(m_c)!}
\frac{d^{m_d}}{(m_d)!}\cdot\frac{(H^1)^{l_1}}{(l_1)!}\cdots\nonumber\\
& &\cdot\frac{(H^r)^{l_r}}{(m_l)!}\frac{c^{l_c}}{(l_c)!}
\frac{d^{l_d}}{(l_d)!}({\rm ln}q)^{\sum_{i=1}^r(m_i+l_i)+m_c+m_d+l_c+l_d}
|\Lambda>\nonumber\\
& &\cdot {\rm tr}\{\pi_{\Lambda_0}[q^{2h_\rho}(H^1)^{m_1}\cdots (H^r)^{m_r}
 d^{m_c}c^{m_d}(H_1)^{l_1}\cdots (H_r)^{l_r}d^{l_c}c^{l_d}]\}\label{c1}
\end{eqnarray}
Clearly $V(\Lambda_0)$ admits a ${\bf Z}$-gradation
\begin{equation}
V(\Lambda_0)=\bigoplus_{s\geq 0}V^{(s)}(\Lambda_0),~~~~V(\Lambda_0)=
\{w\in V(\Lambda_0)|dw=(\tau_0-s)w\}\lable{gradation}
\end{equation}
This means that we can write
\begin{equation}
V(\Lambda_0)\equiv V(\lambda_0,\kappa_0,0)=\bigoplus_{(\lambda_0',\kappa_0,-s)
\in D^+}\bigoplus
_{s\geq 0}\;n_{\lambda_0',s}\,V(\lambda_0',\kappa_0,-s)
\end{equation}
where $n_{\lambda_0',s}$ is the mutliplicity of weight $(\lambda_0',\kappa_0,
-s)$. After some straightforward work we obtain from (\ref{c1})
\begin{equation}
\chi_\Lambda(C^{\Lambda_0})=\sum_{(\lambda_0',\kappa_0,-s)\in D^+}\,
q^{2(\lambda_0',\lambda+\rho_0)}\,\sum_{s=0}^\infty n_{\lambda_0',s}
q^{-2s(\kappa+g)}\label{eigenvalue1}
\end{equation}
which is seen to be absolutely covergent for $|q|\,>\,1$.

We now construct a family of
Casimir invariants. We state our result in the following
\vskip.1in
\noindent{\bf Proposition 2:} Let $\Gamma$ be an operator in (\ref{gamma}).
Then the operators $C^{\Lambda_0}_m$ defined by
\begin{equation}
C^{\Lambda_0}_m=(I\otimes{\rm tr})\{[I\otimes\pi_{\Lambda_0}(q^{2h_\rho})]
\Gamma^m\}\,,~~~~~m\in{\bf Z}^+\label{propos.2}
\end{equation}
are the family of Casimir invariants of $U_q(\hat{\cal G})$. Acting on an
integrable irreducible highest weight $U_q(\hat{\cal G})$-module
$V(\Lambda)$ with highest weight $\Lambda$, the
$C_m^{\Lambda_0}$ take the following eigenvalues
\begin{eqnarray}
\chi_\Lambda(C^{\Lambda_0}_m)=&\sum_{(\lambda+\lambda_0',\kappa+\kappa_0,-s)
\in D^+}&\sum^\infty_{s=0}m_{\lambda_0',s}
\, q^{m(\lambda_0',\lambda_0'+2\lambda+2\rho_0)
-m(\lambda_0,\lambda_0+2\rho_0)-2ms(\kappa+\kappa_0+g)}\nonumber\\
& &\cdot \frac{D_q[(\lambda+\lambda_0',\kappa+\kappa_0,-s)]}
{D_q[(\lambda,\kappa,0)]}\,,~~~~~~m\in{\bf Z}^+\label{proposition2}
\end{eqnarray}
where $m_{\lambda_0',s}$ are multiplicities (see below)
The eigenvalues (\ref{proposition2}) are absolutely covergent for $|q|\,>\,1$.
\vskip.1in
\noindent{\bf Proof:} The statement that $C^{\Lambda_0}_m$ are Casimir
invariants is easy to see: since $\Gamma$ satisfies $\Delta_{\Lambda_0}(a)
\Gamma=\Gamma\Delta_{\Lambda_0}(a)\,~\forall a\in U_q(\hat{\cal G})$, so do its
higher powers; thus by proposition 1, $C^{\Lambda_0}_m$ must be Casimir
inveriants of $U_q(\hat{\cal G})$. We now come to the second part of the
proposition. By (\ref{vv1}) we have
\begin{equation}
\Gamma=(I\otimes\pi_{\Lambda_0})R^TR=(I\otimes \pi_{\Lambda_0})
((v\otimes v)\Delta(v^{-1}))=(v\otimes\pi_{\Lambda_0}(v))\partial
(v^{-1})\label{2}
\end{equation}
where $\partial$ is the algebra homomorphism defined by
\begin{eqnarray}
&&\partial \,:\, U_q(\hat{\cal G})\longrightarrow U_q(\hat{\cal G})\otimes
  {\rm End}V(\Lambda_0)\nonumber\\
&&\partial (v^{-1})=(I\otimes\pi_{\Lambda_0})\Delta (v^{-1})
\end{eqnarray}
We may decompose the tensor product $V(\Lambda)\otimes V(\Lambda_0)$
according to
\begin{equation}
V(\lambda,\kappa,0)\otimes V(\lambda_0,\kappa_0,0)=\bigoplus_{
(\lambda+\lambda_0',\kappa+\kappa_0,-s)\in D^+}
\bigoplus_{s\geq 0}m_{\lambda_0',s}
V(\lambda+\lambda_0',\kappa+\kappa_0,-s)\label{3}
\end{equation}
where $m_{\lambda_0',s}$ are the multiplicities of the modules
$V(\lambda+\lambda_0',\kappa+\kappa_0,-s)$ in the above decomposition.
Note that the sum over $\lambda_0'$ is finite!

Now it follows from (\ref{3}) that on $V(\lambda,\kappa,0)$,
$\Gamma$ in (\ref{2}) takes the value:
\begin{equation}
\alpha_{\lambda_0',s}(\Lambda)=q^{(\lambda_0',\lambda_0'+2\lambda+2\rho_0)
-(\lambda_0,\lambda_0+2\rho_0)-2s(\kappa+\kappa_0+g)}\label{alpha}
\end{equation}
Let $P[\lambda_0',s]$ be the central projections:
\begin{equation}
P[\lambda_0',s] (V(\lambda,\kappa,0)\otimes V(\lambda_0,\kappa_0,0))={V}
(\lambda+\lambda_0',\kappa+\kappa_0,-s)
\end{equation}
With the help of the projection operators, $\Gamma^m$ can be expressed as
\begin{equation}
\Gamma^m=\sum_{(\lambda+\lambda_0',\kappa+\kappa_0,-s)\in D^+}
\sum_{s=0}^\infty\,\alpha^m_{\lambda_0',s}(\Lambda)P[\lambda_0',s]
\end{equation}
Inserting them into (\ref{propos.2}) and noting that $C_m^{\Lambda_0}$ are
Casimir invariants we find
\begin{equation}
\chi_\Lambda(C_m^{\Lambda_0})=\sum_{(\lambda+\lambda_0',\kappa+\kappa_0,
-s)\in D^+}\sum_{s=0}^\infty\,\alpha^m_{\lambda_0',s}(\Lambda)
(I\otimes{\rm tr})\{(I\otimes\pi_{\Lambda_0}(q^{2h_\rho}))P[\lambda_0',s]\}
\end{equation}
which gives, after some effort
\begin{equation}
\chi_\Lambda(C_m^{\Lambda_0})
=\sum_{(\lambda+\lambda_0',\kappa+\kappa_0,-s)\in D^+}
\sum_{s=0}^\infty\,m_{\lambda_0',s}\,
\alpha^m_{\lambda_0',s}(\Lambda)\frac{D_q[(\lambda+\lambda_0',\kappa+\kappa_0,
-s)]}{D_q[(\lambda,\kappa,0)]}\,,~~~m\in{\bf Z}^+\label{eigenvalue2}
\end{equation}
We see using (\ref{alpha}) and (\ref{q-dimension1}) that the r.h.s. of
(\ref{eigenvalue2}) is absolutely covergent for $|q|\,>\,1$.~~~~$\Box$

By comparing (\ref{eigenvalue2}) with (\ref{eigenvalue1}), we arrive at
the interesting identity
\begin{eqnarray}
&&q^{(\lambda_0,\lambda_0+2\rho_0)}\,D_q[(\lambda,\kappa,0)]\,
\sum_{(\lambda_0',\kappa_0,-s)\in D^+}\,
q^{2(\lambda_0',\lambda+\rho_0)}\,\sum_{s=0}^\infty n_{\lambda_0',s}
q^{-2s(\kappa+g)}\nonumber\\
&&~~~~~=\sum_{(\lambda+\lambda_0',\kappa+\kappa_0,-s)\in D^+}
\sum_{s=0}^\infty\,m_{\lambda_0',s}\,q^{(\lambda_0',\lambda_0'+2\lambda
+2\rho_0)-2s(\kappa+\kappa_0+g)}
D_q[(\lambda+\lambda_0',\kappa+\kappa_0,-s)]
\end{eqnarray}
Both side is absolutely covergent for $|q|\,>\,1$.

In summary, we have obtained obviously the Casimir invariants for quantized
affine Lie algebras and computed their eigenvalues for any integrable
irreducible highest weight representation. The eigenvalues
are absolutely covergent for $|q|\,>\,1$.

Finally we remark that we may equivalently work with the coproduct  and
antipode
\begin{eqnarray}
&&\bar{\Delta}(q^h)=q^h\otimes q^h\,,~~~h=h_i,~d\,,~~~i=0,1,\cdots,
r\nonumber\\
&&\bar{\Delta}(e_i)=q^{h_i/2}\otimes e_i+e_i\otimes q^{-h_i/2}\nonumber\\
&&\bar{\Delta}(f_i)=
q^{h_i/2}\otimes f_i+f_i\otimes q^{-h_i/2}\nonumber\\
&&\bar{S}(a)=-q^{-h_\rho}aq^{h_\rho}\,,~~~a=e_i,f_i,h_i,d\label{coproduct2}
\end{eqnarray}
which are obtained by making the interchange $q\leftrightarrow q^{-1}$ in
(\ref{coproduct1}).
Then the universal R-matrix $R$ implements the change $R\leftrightarrow R^T$.
Carrying on the
similar calculations above, we are able to obtain another set of family of
Casimir invariants which are given by the similar formulae above with
$q\leftrightarrow q^{-1}$ and thus are absolutely covergent for $|q|\,<\,1$.
Unfortunately, both sets of invariants appear to diverge in the limit
$|q|\rightarrow 1$.

\vskip.3in
\noindent {\bf Acknowledgements:} Y.Z.Z would like to thank Hoong-Chien
Lee for communication of reference \cite{Lee} and for some comments.
The financial support from the
Australian Research Council is gratefully acknowledged.

\end{document}